\documentclass[aps,prd,amsmath,amssymb,eqsecnum,nofootinbib]{revtex4}

\allowdisplaybreaks[1]
\usepackage{graphicx}

\begin{document}

\title{Quasi-local contribution to the scalar self-force: Non-geodesic Motion}
\author{Adrian C. Ottewill}
\email{adrian.ottewill@ucd.ie}
\author{Barry Wardell}
\email{barry.wardell@ucd.ie}
\affiliation{Complex and Adaptive Systems Laboratory and School of Mathematical Sciences, University College Dublin, Belfield, Dublin 4, Ireland}

\date{\today}

\begin{abstract}
We extend our previous calculation of the quasi-local contribution to the self-force on a scalar particle to general (not necessarily geodesic) motion in a general spacetime. In addition to the general case and the case of a particle at rest in a stationary spacetime, we consider as examples a particle held at rest in Reissner-Nordstr\"{o}m and Kerr-Newman space-times. This allows us to most easily analyse the effect of non-geodesic motion on our previous results and also allows for comparison to existing results for Schwarzschild spacetime. 
\end{abstract}
\maketitle

\section{Introduction}
\label{sec:intro}

In previous work \cite{Ottewill:2007mz} (Paper~I), we calculated the quasi-local contribution to the self-force on a scalar particle in geodesic motion in a general spacetime. We now show how this can be extended to allow for the motion to be non-geodesic. A strong motivation for doing so is that it allows our results to be compared against existing work. Anderson and Hu \cite{Anderson:2003qa} have calculated the coordinate expansion of the function $V(x,x')$ appearing in the Hadamard form of the Green's function, providing a convincing check on the validity of the relations between covariant and coordinate expansions given below. In fact, the comparison allowed a missing factor of 2 in the results of Ref.~\cite{Anderson:2003qa} to be discovered \cite{Anderson:2003qa-e1,Anderson:2003qa-e2}. Anderson and Wiseman \cite{Anderson:2005gb} built on this work to calculate the quasi-local contribution to the scalar self-force for the case of a particle held at rest in the Schwarzschild spacetime. This is useful not only as a check on the current work (in particular, with $Q=0$, for Eqs.~(\ref{eq:staticresult})) but also allows us to correct their result. Finally, Wiseman \cite{Wiseman:2000rm} has shown that the total self-force in this case of a static scalar particle in Schwarzschild spacetime is zero. This could potentially facilitate a study of the usefulness and accuracy of the matched expansion approach \cite{Ottewill:2007mz,Anderson:2005gb} to the calculation of the self-force.

In Sec.~\ref{sec:qlsf-eom}, we extend the general equations of motion calculated in Paper~I to allow for non-geodesic motion. We then show in Sec.~\ref{sec:stationary} how these expressions simplify significantly if the space-time is assumed to be stationary, and that they simplify further under the assumption that the space-time is static. As examples, the cases of a particle at rest in Reissner-Nordstr\"{o}m and Kerr-Newman space-times are considered in Secs.~\ref{sec:rn} and \ref{sec:kn}, respectively. The Reissner-Nordstr\"{o}m result is also used as a check on the existing results of Refs.~\cite{Anderson:2005gb} and \cite{Anderson:2003qa}.

Throughout this paper, we use units in which $G=c=1$ and adopt the sign conventions of \cite{Misner:1974qy}. We denote symmetrization of indices using brackets (e.g. $(\alpha \beta)$) and exclude indices from symmetrization by surrounding them by vertical bars (e.g. $(\alpha | \beta | \gamma)$). Roman letters are used for free indices and Greek letters for indices summed over all space-time dimensions. The Roman letters $i, j, k, l$ are used for indices over spatial dimensions only.

\section{The Quasi-local scalar self-force and equations of motion}
\label{sec:qlsf-eom}
In Paper~I, we calculated an expression for the scalar self-force on a scalar charge $q$ and mass $m$ travelling on a curved background spacetime. However, this expression was only valid provided the particle's path was that of a geodesic of the background spacetime. Fortunately, with some care, this result can be extended relatively easily to allow for non-geodesic particle motion. As the calculation is largely the same as for geodesic motion, we briefly review it here (focusing mostly on the differences caused by the non-geodesicity of the motion) and direct the reader to Paper~I for more extensive coverage.

We begin with our expression from Paper~I (and previously given without some terms in Refs.~\cite{Quinn:2000wa} and \cite{Poisson:2003nc}) for the self-force in terms of a sum of local and non-local parts:
\begin{equation}
\label{eq:FullForce}
f^{a} = q^2 \left( \frac{1}{3} \left( \dot{a}^{a} - a^2 u^{a} \right) + \frac{1}{6} \left( R^{a \beta} u_{\beta} + R_{\beta\gamma} u^\beta u^\gamma u^a \right) +\left(\frac{1}{2}m_{\rm field}^2 - \frac{1}{12} \left( 1 - 6 \xi \right) R \right) u^a + \lim _{\epsilon \rightarrow 0} \int_{-\infty}^{\tau - \epsilon} \nabla^{a} G_{ret} \left( x,x' \right) d\tau '\right)
\end{equation}
where $u^{a}$ is the 4-velocity of the particle, $a$ is the 4-acceleration, $\dot{a} = \frac{\partial{a}}{\partial{\tau}}$ is the derivative of the 4-acceleration with respect to proper time, $G_{ret}(x,x')$ is the retarded scalar Green's function, $m_{\rm field}$ is the field mass and $\xi$ is the coupling to the background scalar curvature.

The first terms are all local to the particle's position and can be computed immediately without posing any real difficulty. Our task is therefore to elucidate the non-local integral term. To this end, we will work with an expression for the self-force which just contains the non-local integral term, with the understanding that the local terms can easily be added back in later if necessary:
\begin{equation}
\label{eq:f-nl}
f^{a}_\mathrm{NL} = \lim _{\epsilon \rightarrow 0} q^2 \int_{-\infty}^{\tau - \epsilon} \nabla^{a} G_{ret} \left( x,x' \right) d\tau '
\end{equation}
Note that in the specific cases considered in Paper~I, we only had geodesic motion ($a^{a}=0$, $\dot{a}=0$) in Ricci-flat spacetimes ($R_{ab}=0$, $R=0$) and without field mass ($m_{\rm field}=0$), so that the local terms were all identically zero. However, for non-geodesic motion, the 4-acceleration will always be non-zero so there will always be at least those terms to be added back in.

As in Paper~I, we will focus only on the the quasi-local contribution to this integral and leave the remaining portion to be computed by other means. The Hadamard form for the retarded Green's function is \cite{Hadamard,Friedlander}:
\begin{equation}
\label{eq:Hadamard}
G_{ret}\left( x,x' \right) = \theta_{-} \left( x,x' \right) \left\lbrace U \left( x,x' \right) \delta \left( \sigma \left( x,x' \right) \right) - V \left( x,x' \right) \theta \left( - \sigma \left( x,x' \right) \right) \right\rbrace 
\end{equation}
where $\theta_{-} \left( x,x' \right)$ is analogous to the Heaviside step-function (i.e. $1$ when $x'$ is in the causal past of $x$, $0$ otherwise), $\delta \left( \sigma\left( x,x' \right) \right)$ is the standard Dirac delta function, $U \left( x,x' \right)$ and $V \left( x,x' \right)$ are symmetric bi-scalars having the benefit that they are regular for $x' \rightarrow x$, and $\sigma \left( x,x' \right)$ is the Synge \cite{Synge,Poisson:2003nc,DeWitt:1965jb} world function. Using this form for the Green's function, the expression for the quasi-local contribution to the scalar self-force simply becomes \cite{Quinn:2000wa}:
\begin{equation}
\label{eq:QLSFInt}
f^{a}_\mathrm{QL} = -q^2 \int_{\tau - \Delta \tau}^{\tau} \nabla^{a} V \left( x,x' \right) d\tau '
\end{equation}
where $\tau - \Delta \tau$ is a \emph{matching point} chosen so that $x(\tau)$ and $x' (\tau')$ are within a convex normal neighborhood  and so that the remainder of the integral in Eq.~(\ref{eq:f-nl}) may be evaluated by other means.

We now expand $V(x,x')$ in two different ways. First, we express it in the form of a covariant Taylor series expansion, i.e. an expansion in increasing powers of the derivative
%\footnote{Since $\sigma$ is a scalar, we can equivalently take either covariant or partial derivatives of it provided we take some care in remaining consistent.} 
of the Synge world function, $\sigma^{a} \equiv \nabla^{a} \sigma = O(\sigma^{1/2})$:
\begin{equation}
\label{eq:Vt}
V \left( x,x' \right) = \sum _{p=0}^{\infty} \frac{ \left( -1 \right) ^p }{p!} v_{\alpha_1 \dots \alpha_p} (x) \sigma ^{\alpha_1} \left( x,x' \right) \dots \sigma ^ {\alpha_p} \left( x,x' \right)
\end{equation}
where explicit expressions for the $v_{a_1 \dots a_p} (x)$, up to $O \left( \sigma^{5/2} \right)$ are given in Paper~I for massless fields in vacuum spacetimes. Paper~I also describes how they may be obtained for massive fields in general spacetimes from the results Ref.~\cite{Decanini:2005gt}.

\begin{figure}
\includegraphics[width=6cm]{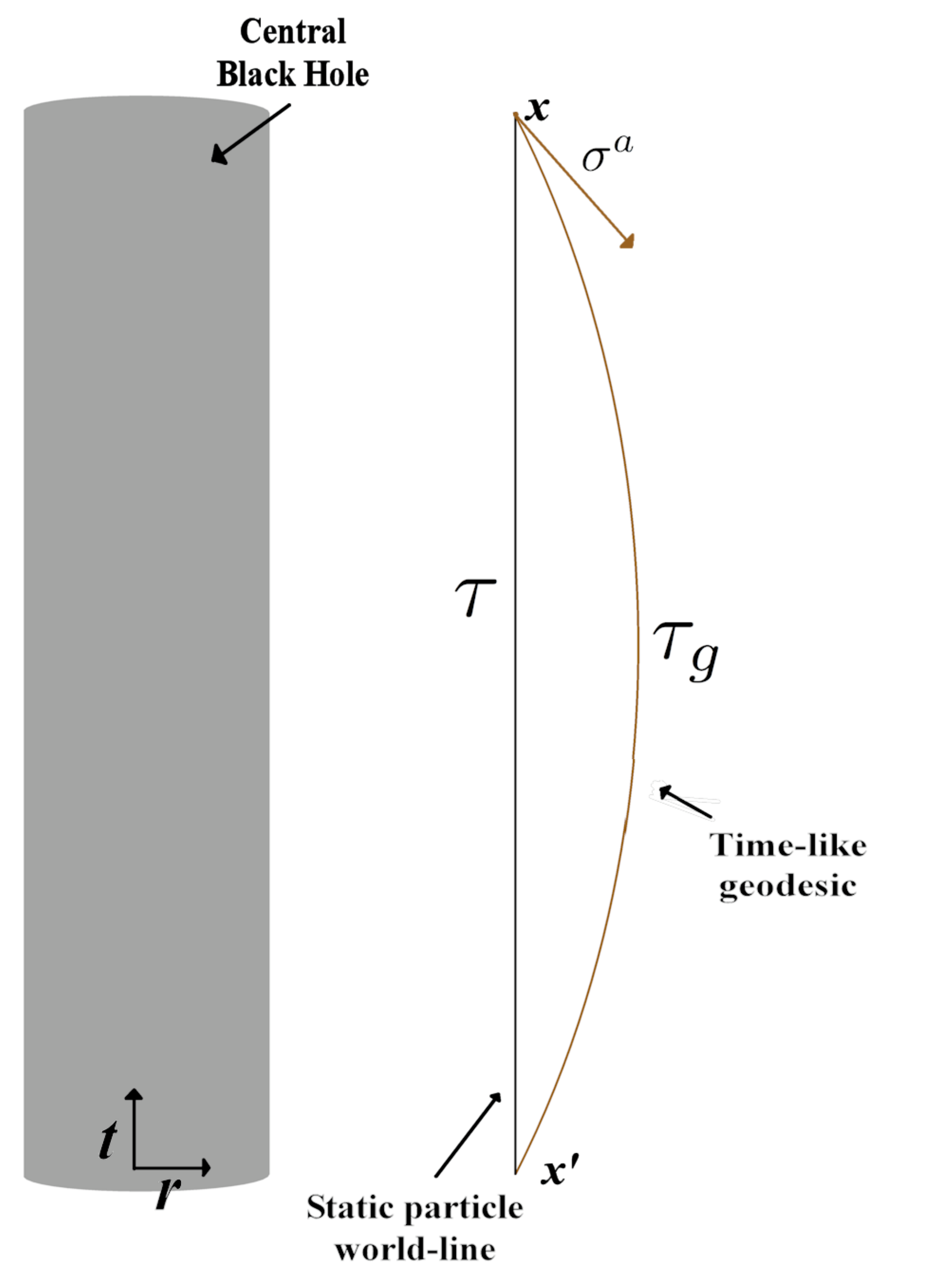}
\caption{The world-line of a static particle ($x^i=\mathrm{constant}$) in Reissner-Nordstr\"{o}m space-time is not a geodesic, so the geodesic proper time, $\tau_g$, and the particle proper time, $\tau$, are not the same. The integral in Eq.~(\ref{eq:QLSFInt}) is along the particle world-line, while $\sigma^a$ is along the time-like geodesic.}
\label{fig:non-geodesic}
\end{figure}

In Paper~I, we were able to use this expression for $V(x,x')$ to compute the self-force for geodesic motion. Unfortunately, things are less straightforward when non-geodesic motion is allowed for. The problem arises as a result of the presence of $\sigma^a$ in this expression. As demonstrated in Fig.~(\ref{fig:non-geodesic}), it encodes the proper time, $\tau_g$, along a geodesic of the background space-time through $\sigma^a\sigma_a=- \tau_g{}^2$. However, the integral in Eq.~(\ref{eq:QLSFInt}) is along the world line of the particle.  In the previous case of geodesic motion, this was not a problem as in that case $\tau_g$ is a natural parameter along the world line. For non-geodesic motion, this is no longer the case. To proceed with the calculation using this expansion of $V(x,x')$ would require us to first express the geodesic proper time $\tau_g$ in terms of the integration variable, i.e. the particle's proper time $\tau$. This is a non-trivial task for general motions in general space-times.
An easier resolution of this problem comes from expressing $V(x,x')$ in a second form, as a non-covariant Taylor series expansion in the coordinate separation of the points, $\Delta x^a$:
\begin{equation}
\label{eq:Vc}
V \left( x,x' \right) = \sum _{p=0}^{\infty} \frac{ \left( -1 \right) ^p }{p!} \hat{v}_{\alpha_1 \dots \alpha_p} (x) \Delta x^{\alpha_1} \dots \Delta x^{\alpha_p}
\end{equation}
where $\Delta x^{\alpha} = x^\alpha - x^{\alpha '}$ and the quantities $\hat{v}_{\alpha_1 \dots \alpha_p} (x)$ may be expressed in terms of combinations of the (known) quantities $v_{\alpha_1 \dots \alpha_p} (x) $ as follows.

First, the world function $\sigma \left(x,x'\right)$ may be expressed in terms of an expansion in powers of the coordinate separation of $x$ and $x'$, $\Delta x^\alpha$:
\begin{equation}
\label{eq:sigma-coord}
 \sigma = \frac{1}{2} g_{\alpha \beta} \Delta x^\alpha \Delta x^\beta
	 + A_{\alpha \beta \gamma} \Delta x^\alpha \Delta x^\beta \Delta x^\gamma
	 + B_{\alpha \beta \gamma \delta} \Delta x^\alpha \Delta x^\beta \Delta x^\gamma \Delta x^\delta
	 + C_{\alpha \beta \gamma \delta \epsilon} \Delta x^\alpha \Delta x^\beta \Delta x^\gamma \Delta x^\delta  \Delta x^\epsilon
	 + \ldots
\end{equation}
where the coefficients $A_{\alpha \beta \gamma}$, $B_{\alpha \beta \gamma \delta}$, $C_{\alpha \beta \gamma \delta \epsilon}$, \dots are to be determined. In order to determine these coefficients, we note that Eq.~(\ref{eq:sigma-coord}) implies the coordinate expansion of the derivative of $\sigma$ is:
\begin{equation}
\label{eq:sigmaderiv-coord}
 \sigma_a = g_{a \alpha} \Delta x^\alpha
	 + (\frac12 g_{\alpha \beta ,a} + 3 A_{a \alpha \beta} )\Delta x^\alpha \Delta x^\beta
	 + (A_{\alpha \beta \gamma ,a} + 4 B_{a \alpha \beta \gamma}) \Delta x^\alpha \Delta x^\beta \Delta x^\gamma
	 + (B_{\alpha \beta \gamma \delta ,a} + 5 C_{a \alpha \beta \gamma \delta} )\Delta x^\alpha \Delta x^\beta \Delta x^\gamma \Delta x^\delta
	 + \ldots .
\end{equation}
Substituting expansions (\ref{eq:sigma-coord}) and (\ref{eq:sigmaderiv-coord}) into the defining relationship
\begin{equation}
\label{eq:sigma-to-sigmaderiv}
 2 \sigma = \sigma^\alpha \sigma_\alpha
\end{equation}
and equating powers of $\Delta x^\alpha$, we get expressions for each coefficient in terms of the lower order coefficients. The lowest terms are given by:
\begin{subequations}
\begin{eqnarray}
 A_{a b c} 	&=& -\frac{1}{4}g_{(a b ,c)} \\
 B_{a b c d}	&=& -\frac{1}{3} \left( A_{(a b c ,d)} +  g^{\alpha \beta} \left(\frac{1}{8}g_{(a b ,|\alpha|} g_{c d) ,\beta}
			+ \frac{3}{2} g_{(a b ,|\alpha|} A_{|\beta| c d)}
			+ \frac{9}{2} A_{\alpha (a b} A_{|\beta| c d)}\right) \right) \\
 C_{a b c d e}	&=& -\frac{1}{4} \left( B_{(a b c d ,e)} +  g^{\alpha \beta} \left(12 A_{\alpha (a b} B_{|\beta| c d e)}
			+ 3 A_{\alpha (a b} A_{c d e) ,\beta}
			+ 2 g_{(a b |,\alpha|} B_{|\beta| c d e)}
			+ \frac{1}{2} A_{(a b c |,\alpha|} g_{d e) ,\beta}\right)\right)
\end{eqnarray}
\end{subequations}
This procedure may be easily extended to higher orders using a computer algebra package.

Next, we obtain a relation between the known coefficients of the covariant expansion, Eq.~(\ref{eq:Vt}), and those of the coordinate expansion, Eq.~(\ref{eq:Vc}), by first substituting Eq.~(\ref{eq:sigmaderiv-coord}) into Eq.~(\ref{eq:Vt}) and then equating the two expansions. In this way we find the following expressions for the $\hat{v}_{\alpha_1 \dots \alpha_p} (x)$ in terms of the $v_{\alpha_1 \dots \alpha_p} (x)$:
\begin{subequations}
\begin{eqnarray}
\hat{v} 	&=& v\\
\hat{v}_a 	&=& v_{a}\\
\label{eq:vab-hat}
\hat{v}_{ab} 	&=& v_{ab} + v_{\alpha} \Gamma^{\alpha}_{ab}\\
\hat{v}_{abc}	&=& v_{abc} + 3 v_{(a |\alpha|} \Gamma_{b c)}^{\alpha} 
		+ 6v_{\alpha} \left( A_{(a b c)}^{\phantom{(a b c)},\alpha} + 4 B^{\alpha}_{\phantom{\alpha} a b c}\right)\\
\hat{v}_{abcd}	&=& v_{abcd} + 6 v_{\alpha (a b} \Gamma_{c d)}^{\alpha}
		+ 12 v_{\alpha \beta} \left(\frac{1}{4}\Gamma_{(a b}^{\alpha}\Gamma_{c d)}^{\beta}+2 g_{(a}^{\phantom{(a}\alpha}A_{b c d)}^{\phantom{b c d)},\beta}+ 8 g_{(a}^{\phantom{(a}\alpha} B^{\beta}_{\phantom{\beta} b c d)}\right)
		\nonumber \\
		& & -24 v_{\alpha} \left( B_{(a b c d)}^{\phantom{(a b c d)},\alpha}
			+5C^{\alpha}_{\phantom{\alpha} (a b c d)}\right)\\
\label{eq:vhat5}
\hat{v}_{abcde}	&=& \frac{1}{2} \hat{v}_{,(a b c d e)} - \frac{5}{2} \hat{v}_{(a b , c d e)} + \frac{5}{2} \hat{v}_{(a b c d , e)} 
\end{eqnarray}
\end{subequations}
where the $\Gamma_{a b}^{\alpha}$ are the Christoffel symbols of the second kind.

Although $\hat{v}_{abcde}$ could alternatively be given in terms of the $v_{\alpha_1 \dots \alpha_p} (x)$, we have found that the expression we give proves easier to work with. It is found by taking five symmetrized partial derivatives of the equation
\begin{equation}
\label{eq:vsymmetry}
 V(x,x') = V(x',x)
\end{equation}
and then taking taking the coincidence limit $x' \rightarrow x$.  It is a special case of the general result that follows from taking any number of symmetrized partial derivatives:
\begin{equation}
\label{eq:vcompsym}
\hat{v}_{a_1a_2\dots a_p}	=  \frac{1}{2}\sum\limits_{k=0}^{p-1} \binom{p}{k} (-1)^k \hat{v}_{(a_1 a_2 \dots a_{k} , a _{k+1} \dots  a_{p})}  \quad \text{for } p \text{ odd}
\end{equation}
Applying this identity recursively, we can re-express this with all odd lower order coefficients eliminated:
\begin{equation}
\label{eq:vcompsym2}
\hat{v}_{a_1a_2\dots a_p}	=  \sum\limits_{\substack{k=0\\k~\text{even}}}^{p-1}\binom{p}{k} \frac{2(2^{p-k+1}-1)}{p-k+1} B_{p-k+1} \hat{v}_{(a_1 a_2 \dots a_{k} , a _{k+1} \dots  a_{p})} \quad \text{for } p \text{ odd}
\end{equation}
where the $B_n$ are the Bernoulli numbers \cite{GradRyz}. Thus, these identities determine all odd coefficients in terms of derivatives of lower order even coefficients. 

Now, we simply substitute expansion (\ref{eq:Vc}) into Eq.~(\ref{eq:QLSFInt}) and, since $V(x,x')$ is a scalar, take a partial rather than covariant derivative to get an easily evaluated expression for the self-force (in this case, it is most natural to work with an expression for the self-force in covariant rather than contravariant form):
\begin{eqnarray}
\label{eq:DxExpandedIntegral}
f_{a}^{\rm QL} &=& -q^2 \int_{\tau - \Delta \tau}^{\tau} \Big[
	\hat{v}_{,a} - \hat{v}_{a}
	- \left( \hat{v}_{\alpha ,a} - \hat{v}_{\alpha a} \right) \Delta x^\alpha
	+ \frac{1}{2} \left( \hat{v}_{\alpha \beta ,a} - \hat{v}_{\alpha \beta a} \right) \Delta x^\alpha \Delta x^\beta \nonumber \\
	& &
	- \frac{1}{3!} \left( \hat{v}_{\alpha \beta \gamma ,a} - \hat{v}_{\alpha \beta \gamma a} \right) \Delta x^\alpha \Delta x^\beta \Delta x^\gamma
	+ \frac{1}{4!} \left( \hat{v}_{\alpha \beta \gamma \delta ,a} - \hat{v}_{\alpha \beta \gamma \delta a} \right) \Delta x^\alpha \Delta x^\beta \Delta x^\gamma \Delta x^\delta
	+ O\left( \Delta x^{5} \right)
	\Big] d\tau '
\end{eqnarray}

Finally, we obtain the equations of motion by projecting orthogonal and perpendicular to the particle 4-velocity:
\begin{eqnarray}
\label{eq:ma}
ma^{a} = P^{a \beta} f_{\beta}\\
\label{eq:dmdtau}
\frac{dm}{d\tau} = -f_{\alpha} u^{\alpha}
\end{eqnarray}
where
\begin{equation}
\label{eq:Projection}
P^{a b} = g^{a b} + u^{a} u^{b}
\end{equation}
is the projection orthogonal to $u^\alpha$.

This expression may be evaluated for a specific particle path in a specific spacetime by writing the coordinate separations $\Delta x^\alpha$ in terms of the particle proper time separation $\left(\tau - \tau'\right)$. Specific examples of such an evaluation are given in the following sections.

\section{Particle ``at rest'' in a stationary spacetime}
\label{sec:stationary}
The expression for the quasi-local contribution to the self-force given in Eq.~(\ref{eq:DxExpandedIntegral}) and the subsequent equations of motion take on a significantly simple form if the space-time is assumed to be stationary and if we assume the spatial coordinate of the particle to be fixed. Stationarity allows us to introduce coordinates $(t,x^i)$ such that the metric tensor components, $g_{a b}$ are all independent of the time coordinate. In addition, for a particle ``at rest", $x^i = \mathrm{constant}$, and only the time component of the contravariant 4-velocity, $u^t$, is non-zero. These conditions hold, of course, for the case of a particle held at rest in Schwarzschild spacetime which has already received much attention in the literature \cite{Anderson:2005gb,Wiseman:2000rm}.

Since the spatial coordinates of the particle are held fixed, the points $x$ and $x'$ are now only separated in the time direction. Furthermore, in this case it is straightforward to relate the time coordinate to the proper time along the particle's world line. As a result we can rewrite the coordinate separation of the points $x$ and $x'$ in terms of the proper time:
\begin{equation}
 \Delta x^t = \Delta t = u^t (\tau - \tau ')  = \frac{1}{\sqrt{-g_{tt}}} (\tau - \tau ')  , \qquad \Delta x^i =0.
\end{equation}
Substituting this into Eq.~(\ref{eq:DxExpandedIntegral}) and performing the straightforward integral of powers of $(\tau-\tau')$ gives
\begin{eqnarray}
\label{eq:force-integrated}
f_{a}^{\rm QL} &=& -q^2\Big[
	\left( \hat{v}_{,a} - \hat{v}_{a} \right) \Delta \tau
	- \frac{1}{2!} \left( \hat{v}_{t ,a} - \hat{v}_{t a} \right) u^t \Delta \tau ^2
	+ \frac{1}{3!} \left( \hat{v}_{t t ,a} - \hat{v}_{t t a} \right) (u^t)^2 \Delta \tau ^3 \nonumber \\
	& &
	- \frac{1}{4!} \left( \hat{v}_{t t t ,a} - \hat{v}_{t t t a} \right) (u^t)^3 \Delta \tau ^4
	+ \frac{1}{5!} \left( \hat{v}_{t t t t ,a} - \hat{v}_{t t t t a} \right) (u^t)^4 \Delta \tau ^5
	+ O\left( \Delta \tau ^{6} \right)
	\Big]
\end{eqnarray}

We can now use Eq.~(\ref{eq:vhat5}) for $\hat{v}_{a b c d e}$, along with the analogous equations for  $\hat{v}_{a}$ and $\hat{v}_{a b c}$,
\begin{eqnarray}
 \hat{v}_{a} &=& \frac{1}{2} \hat{v}_{,a}\\
 \hat{v}_{a b c} &=& -\frac{1}{4} \hat{v}_{,(a b c)} + \frac{3}{2} \hat{v}_{(a b , c)} 
\end{eqnarray}
to eliminate several terms in this expression. While this did not previously prove particularly beneficial in the general case, the fact that partial derivatives with respect to $t$ of these fundamentally geometric objects vanish in a stationary spacetime means that many of the extra terms introduced by the substitution will also vanish. Indeed it is easy to see from Eq.~(\ref{eq:vcompsym}) that (in this case of purely time separated points) any term of order $2k+1$ can be related to the order $2k$ term:
\begin{equation}
\hat{v}_{\underbrace{\scriptstyle tt\dots tt}_{2k}b} = \frac{1}{2} \hat{v}_{{\underbrace{\scriptstyle{tt\dots tt}}_{2k}},b} 
\end{equation}
As a result, Eq.~(\ref{eq:force-integrated}) may be taken to arbitrary order to give:
\begin{eqnarray}
\label{eq:force-integrated-simplified}
f_{a}^{\rm QL} &=& -q^2 \Big[
	\frac{1}{2} \sum_{k=0}^{\infty} \frac{1}{(2k+1)!} \hat{v}_{\underbrace{\scriptstyle tt\dots tt}_{2k}, a} \Delta \tau^{2k+1} (u^t)^{2k}
	+ \sum_{k=1}^{\infty} \frac{1}{(2k)!} \hat{v}_{\underbrace{\scriptstyle tt\dots tt}_{2k-1} a} \Delta \tau^{2k} (u^t)^{2k-1}
	\Big]
\end{eqnarray}

To proceed further, we benefit from differentiating between the time and spatial components:
\begin{eqnarray}
\label{eq:force-integrated-covariant-time}
f_{t}^{\rm QL} &=& -q^2 \Big[
	\sum_{k=1}^{\infty} \frac{1}{(2k)!} \hat{v}_{\underbrace{\scriptstyle tt\dots tt}_{2k}} \Delta \tau^{2k} (u^t)^{2k-1}
	\Big]\\
\label{eq:force-integrated-covariant-spatial}
f_{i}^{\rm QL} &=& -q^2 \Big[
	\frac{1}{2} \sum_{k=0}^{\infty} \frac{1}{(2k+1)!} \hat{v}_{\underbrace{\scriptstyle tt\dots tt}_{2k}, i} \Delta \tau^{2k+1} (u^t)^{2k}
	+ \sum_{k=1}^{\infty} \frac{1}{(2k)!} \hat{v}_{\underbrace{\scriptstyle tt\dots tt}_{2k-1} i} \Delta \tau^{2k} (u^t)^{2k-1}
	\Big]
\end{eqnarray}

From this, it is clear that the $t$ component of the quasi-local self-force only appears at even orders in $\Delta \tau$ while the spatial components may appear at both even and odd orders. Note, however, that for a static spacetime - i.e. imposing time reversal invariance in addition to stationarity - the second term in Eq.~(\ref{eq:force-integrated-covariant-spatial}) will vanish since each of the $\hat{v}_{t\dots t i}$ must be zero in that case (as we have an odd number of $t$'s).  As a result, in a static space-time, the spatial component of the quasi-local self-force will only appear at odd orders in $\Delta \tau$.

By Eq.~(\ref{eq:dmdtau}) and the fact that only the time component of the contravariant 4-velocity is non-zero, we can now rewrite the rate of change of the particle's mass as:

\begin{eqnarray}
\label{eq:dmdtau-stationary-spacetime}
 \frac{dm}{d\tau} &=& -f_{\alpha} u^{\alpha} = - f_t u^t \nonumber \\
	&=& q^2
	\sum_{k=1}^{\infty} \frac{1}{(2k)!} \hat{v}_{\underbrace{\scriptstyle tt\dots tt}_{2k}} \Delta \tau^{2k} (u^t)^{2k}
\end{eqnarray}

Similarly, by Eq.~(\ref{eq:ma}) we can now write the time and spatial components of the mass times 4-acceleration:
\begin{subequations}
\label{eq:ma-stationary-spacetime}
\begin{eqnarray}
ma^{t}_{\rm QL} &=& -\frac{g_{ti}g^{ti}}{g_{tt}} f_t + g^{ti} f_i \\
ma^{i}_{\rm QL} &=& g^{ti} f_{t} + g^{i j} f_{j}
\end{eqnarray}
\end{subequations}
where $f_t$ and $f_i$ are as given in Eqs.~(\ref{eq:force-integrated-covariant-time}) and (\ref{eq:force-integrated-covariant-spatial}).

Again, it is interesting to note the effect of imposing that the spacetime be static. For a static spacetime, the metric components odd in $t$ vanish, so our result simplifies to:
\begin{subequations}
\label{eq:ma-static-spacetime}
\begin{eqnarray}
ma^{t}_{\rm QL} &=& 0 \\
ma^{i}_{\rm QL} &=& g^{i j} f_{j} =  -q^2 g^{i j} \frac{1}{2} \sum_{k=0}^{\infty} \frac{1}{(2k+1)!} \hat{v}_{\underbrace{\scriptstyle tt\dots tt}_{2k}, j} \Delta \tau^{2k+1} (u^t)^{2k}
\end{eqnarray}
\end{subequations}

\section{Particle at rest in Reissner-Nordstr\"{o}m spacetime}
\label{sec:rn}
We now look at a specific example, where we calculate the quasi-local contribution to the scalar self-force on a static particle (i.e. held at rest) in Reissner-Nordstr\"{o}m spacetime. We chose this spacetime as while, through its spherical symmetry, it retains much of the simplicity of Schwarzschild spacetime, it also illustrates better the difference non-geodesic motion makes. This is because its non-vanishing Ricci tensor means that the self-force appears at lower order than was seen in Paper~I, where we concentrated on vacuum spacetimes. We therefore have more orders in which to see the difference between using the covariant (Eq.~(\ref{eq:Vt})) and coordinate (Eq.~(\ref{eq:Vc})) expansions for $V(x,x')$.

The Reissner-Nordstr\"{o}m spacetime has line element
\begin{equation}
\label{eq:RNMetric}
ds^2 = \left( 1-\frac{2M}{r} + \frac{Q^2}{r^2}\right)^{-1} dr^2 + r^2 \left(d\theta ^2 + \sin ^2 \theta d\phi ^2 \right) - \left( 1-\frac{2M}{r} + \frac{Q^2}{r^2}\right) dt^2
\end{equation}

Since this is an example of a particle at rest in a static spacetime, we may use Eqs.~(\ref{eq:dmdtau-stationary-spacetime}) and (\ref{eq:ma-static-spacetime}) in order to calculate the equations of motion. Computing and substituting in the expressions for the relevant $\hat{v}_{t \dots t} (x)$ in Reissner-Nordstr\"{o}m spacetime, along with the time component of the 4-velocity, 
\begin{equation}
u^t = \left(1-\frac{2M}{r}+\frac{Q^2}{r^2}\right)^{-1/2}
\end{equation}
we arrive at our result for the quasi-local contribution to the equations of motion of a scalar particle held at rest in Reissner-Nordstr\"{o}m:
\begin{subequations}
\label{eq:staticresult}
\begin{eqnarray}
ma_{\rm QL}^{r} &=& -q^2\Big[\frac{Q^2 \left(Q^2-2 M r+r^2\right)}{60 r^{11}} \left(5 Q^2-8 M r+3 r^2\right) \Delta\tau^3
	+\frac{1}{6720 r^{15}}\big(1568 Q^8-7154 M Q^6 r\nonumber\\
	& & ~~~~~~ +10647 M^2 Q^4 r^2 +2604 Q^6 r^2-5265 M^3 Q^2 r^3-7424 M Q^4 r^3+198 M^4 r^4+5147 M^2 Q^2 r^4\nonumber\\
	& & ~~~~~~+1236 Q^4 r^4-171 M^3 r^5 -1566 M Q^2 r^5+36 M^2 r^6+144 Q^2 r^6\big) \Delta\tau^5 + O(\Delta\tau^7)\Big]\\
ma_{\rm QL}^{\theta} &=& 0\\
ma_{\rm QL}^{\phi} &=& 0\\
ma_{\rm QL}^{t} &=& 0\\
\frac{dm}{d\tau} &=& -q^2\Big[\frac{Q^2 \left(Q^2-2 M r+r^2\right) }{20 r^8}\Delta\tau^2
	+\frac{1}{1344 r^{12}}\big(196 Q^6-588 M Q^4 r+429 M^2 Q^2 r^2+204 Q^4 r^2-18 M^3 r^3 \nonumber \\
	& & ~~~~~~ -268 M Q^2 r^3+9 M^2 r^4+36 Q^2 r^4\big) \Delta\tau^4 + O(\Delta\tau^6)\Big]
\end{eqnarray}
\end{subequations}

In the limit $Q \rightarrow 0$, this reduces to the Schwarzschild case, which may be compared to Ref.~\cite{Anderson:2005gb}. However, Anderson and Wiseman use an incorrect expression to relate $\Delta t$ to $\left(\tau-\tau'\right)$ and also use the results of Ref.~\cite{Anderson:2003qa} prior to the corrections given in the subsequent errata \cite{Anderson:2003qa-e1,Anderson:2003qa-e2}. Once these two issues are corrected, we find our results are in exact agreement.

An alternative approach in the special case of a static, spherically symmetric space-time is to use the Hadamard-WKB approach developed in Ref.~\cite{Anderson:2003qa} to calculate the coordinate expansion of the retarded Green's function, i.e.
\begin{equation}
\label{eq:WKBGreen}
V\left( x,x' \right) = \sum_{i,j,k=0}^{\infty} \tilde v_{ijk} \left( t - t' \right)^{2i} \left( \cos \gamma - 1 \right)^j \left(r - r'\right)^k .
\end{equation}
where the coordinate $\gamma$ is the angle on the 2-sphere between $x$ and $x'$.

Upon doing so for Reissner-Nordstr\"{o}m spacetime, we find that the coefficients of the expansion of $V(x,x')$ relevant to the current static particle calculation are\footnote{Note here that our definition of the retarded Green's function differs from that of Ref.~\cite{Anderson:2003qa} (including the subsequent errata \cite{Anderson:2003qa-e1,Anderson:2003qa-e2}) by a factor of 2 in the term involving $v(x,x')$. The coefficients given here are consistent with the definition of the retarded Green's function given in Eq.~(\ref{eq:Hadamard}).}:
\begin{subequations}
\begin{equation}
 \tilde v_{000}=\tilde v_{001}=0
\end{equation}
and
\begin{eqnarray}
 \tilde v_{100}&=&-\frac{Q^2 \left(Q^2-2 M r+r^2\right)^2 t^2}{20 r^{10}}\\
 \tilde v_{101}&=&-\frac{Q^2 \left(Q^2-2 M r+r^2\right) \left(5 Q^2-8 M r+3 r^2\right)}{20 r^{11}}\\
 \tilde v_{200}&=&-\frac{\left(Q^2-2 M r+r^2\right)^2}{1344 r^{16}} \Big[196 Q^6+12 r (17 r-49 M) Q^4+r^2
   \left(429 M^2-268 r M+36 r^2\right) Q^2\nonumber \\
   && ~~~ +9 M^2 r^3 (r-2 M)\Big]\\
 \tilde v_{201}&=&-\frac{\left(Q^2-2 M r+r^2\right)}{1344 r^{17}} \Big[1568 Q^8+14 Q^6 r (-511 M+186 r)+9 M^2 r^4 \left(22 M^2-19 M r+4 r^2\right) \nonumber \\
   && ~~~ +Q^4 r^2 \left(10647 M^2-7424 M r+1236 r^2\right)+Q^2 r^3 \left(-5265 M^3+5147 M^2 r-1566 M r^2+144 r^3\right)\Big]
\end{eqnarray}
\end{subequations}

Additionally, for a particle held at rest in Reissner-Nordstr\"{o}m, we can immediately relate the coordinate point separations $\Delta x^\alpha$ to $\tau - \tau'$,
\begin{equation}
\label{eq:coord-to-proper}
	\Delta r = 0,~ ~ ~\Delta \theta =0,~ ~ ~\Delta \phi =0,~ ~ ~
	\Delta t =\left(1-\frac{2M}{r}+\frac{Q^2}{r^2}\right)^{-1/2} \left({\tau}-{\tau}'\right)
\end{equation}

Taking the partial derivative of Eq.~(\ref{eq:WKBGreen}), relating the coordinate separations to proper time separations using Eqs.~(\ref{eq:coord-to-proper}) and performing the integral over $\tau'$, we arrive at a result which is in exact agreement with Eqs.~(\ref{eq:staticresult}).

\section{Particle at rest in Kerr-Newman spacetime}
\label{sec:kn}
To illustrate the flexibility of the approach presented in this paper, we will now investigate the case of a particle in Kerr-Newman spacetime at rest relative to an observer at spatial infinity. We choose Kerr-Newman for analogous reasons to those of Sec.~\ref{sec:rn} - the non-vanishing Ricci tensor means that the effects of the non-geodesicity of the motion are more apparent. While it is not clear whether the WKB approach of Ref.~\cite{Anderson:2003qa} could be extended to such a spacetime, the method presented here is easily adapted to any spacetime, including Kerr-Newman.

In Boyer-Linquist coordinates, the Kerr-Newman metric is \cite{Boyer-Lindquist}:
\begin{equation}
\label{eq:KerrNewmanMetric}
ds^2 = - \frac{\Delta - a^2 + z^2}{\rho^2} dt^2 
	+ \frac{2(\Delta-r^2-a^2)(a^2-z^2)}{a \rho^2}dtd\phi
	+ \frac{\rho^2}{\Delta}dr^2
	+ \rho^2  d\theta^2
	+ \frac{a^2-z^2}{a^2 \rho^2} \left( (a^2+r^2)^2 - \Delta (a^2-z^2) \right) d\phi^2
\end{equation}
where $\Delta=r^2-2Mr+a^2+Q^2$, $\rho^2 = r^2 + a^2 \cos^2 \theta$ and $z=a \cos \theta$. Here, $Q$ is the charge, $M$ is the mass and $a$ is the angular momentum per unit mass of the black hole.

For the case of a particle in such a spacetime at rest relative to an observer at spatial infinity, the condition on the motion is that the spatial components of the 4-velocity vanish, i.e. $u^i=0$. Additionally, the time component of the 4-velocity may be written as:
\begin{equation}
\label{eq:coord-to-proper-kn}
	u^t =\sqrt{\frac{\rho^2}{\Delta - a^2 + z^2}}
\end{equation}

Now, since Kerr-Newman is an example of a stationary spacetime, we may use Eqs.~(\ref{eq:dmdtau-stationary-spacetime}) and(\ref{eq:ma-stationary-spacetime}) in order to calculate the equations of motion:
\begin{subequations}
\begin{align}
&\frac{dm}{d\tau} = -\frac{q^2 Q^2}{20 \Delta  \rho ^8}\left(6 a^4+\left\{Q^2+r [-2 M+r]\right\}^2-4 \left\{Q^2+r [-2 M+r]\right\} z^2+z^4+6 a^2 \left\{Q^2-2 M r+r^2-z^2\right\}\right) \Delta\tau^2\notag \\
&\hspace{0.6em}
\begin{aligned}
&\hspace{-0.3em}	- \frac{q^2}{1344 \Delta ^2 \rho ^{14}}\boldsymbol{\Big(} 28 Q^{10} \left\{7 r^2+z^2\right\}-4 Q^8 \left\{[343 M-149 r] r^3+r [7 M+234 r] z^2+131 z^4\right\} +Q^6 \left\{r^4 \left[3565 M^2 \right. \right. \\
&	\left.\left.- 3044 M r+640 r^2\right]-2 r^2 \left[215 M^2-2654 M r+1076 r^2\right] z^2+\left[37 M^2+2304 M r-1100 r^2\right] z^4+684 z^6\right\}\\
&	- Q^4 \left\{[2 M-r] r^5 \left[2043 M^2-1532 M r+276 r^2\right]+r^3 \left[-1124 M^3+9971 M^2 r-7516 M r^2+1440 r^3\right] z^2 \right.\\
&	\left.+ r \left[166 M^3+3119 M^2 r-3140 M r^2+864 r^3\right] z^4+\left[137 M^2+2428 M r-1008 r^2\right] z^6-156 z^8\right\} \\
&	+ Q^2 \left\{r^6 [-2 M+r]^2 \left[447 M^2-268 M r+36 r^2\right]-4 [2 M-r] r^4 \left[85 M^3-739 M^2 r+447 M r^2-63 r^3\right] z^2\right. \\
&	+ 4 r^2 \left[55 M^4+371 M^3 r-545 M^2 r^2+327 M r^3-72 r^4\right] z^4+4 r \left[109 M^3+516 M^2 r-427 M r^2+72 r^3\right] z^6 \\
&	\left.+ 3 \left[27 M^2-104 M r+84 r^2\right] z^8-36 z^{10}\right\}-9 M^2 \left\{[2 M-r]^3 r^3+9 r^2 [-2 M+r]^2 z^2+9 [2 M-r] r z^4+z^6\right\} \rho ^4\\
&	+ 12 a^6 \rho ^2 \left\{154 Q^4+15 M^2 \rho ^2+30 Q^2 \left[-7 M r+2 \rho ^2\right]\right\} +6 a^4 \left\{42 Q^6 \left[13 r^2+9 z^2\right] -8 Q^4 \left[(224 M-89 r) r^3 \right. \right. \\
&	\left.+ 5 (28 M-9 r) r z^2+44 z^4\right]-45 M^2 \left[2 M r-r^2+z^2\right] \rho ^4 +Q^2 \left[M^2 \left(1459 r^4+846 r^2 z^2+59 z^4\right) \right. \\
&	\left.\left.- 20 M r \left(53 r^2-24 z^2\right) \rho ^2+180 \left(r^2-z^2\right) \rho ^4\right]\right\} +4 a^2 \left\{7 Q^8 \left[59 r^2+23 z^2\right] -Q^6 \left[(2135 M-906 r) r^3\right.\right.\\
&	\left. + r (623 M+288 r) z^2+690 z^4\right] + Q^4 \left[r^4 \left(3639 M^2-2994 M r+601 r^2\right) +r^2 \left(670 M^2+1236 M r-543 r^2\right) z^2 \right. \\
&	\left.+ \left(55 M^2+2214 M r-1083 r^2\right) z^4+61 z^6\right] +27 M^2 \left[r^2 (-2 M+r)^2-3 r (-2 M+r) z^2+z^4\right] \rho ^4\\
&	- Q^2 \left[4 M^3 r \left(524 r^4+61 r^2 z^2+41 z^4\right)+M^2 \left(-2446 r^6+1251 r^4 z^2+1776 r^2 z^4+95 z^6\right) \right. \\
&	\left. \left. + 15 M r \left(61 r^4-118 r^2 z^2+3 z^4\right) \rho ^2-108 \left(r^4-3 r^2 z^2+z^4\right) \rho ^4\right]\right\}\boldsymbol{\Big)} \Delta\tau^4 + O(\Delta\tau^6)
\end{aligned}
\end{align}
\begin{align}
&ma^t=-\frac{q^2 \left(Q^2-2 M r\right) \left(a^2-z^2\right) }{75600 \Delta ^{5/2} \rho ^{17}} \boldsymbol{\Bigg[} 
	11340 Q^2 \rho^8 \Delta \left(2 a^2+Q^2-2 M r+r^2-z^2\right)\Delta\tau^2 \notag \\
&\hspace{0.6em}
\begin{aligned}
&\hspace{-0.3em}	+ \boldsymbol{\Big(}Q^8 \left\{23807 r^4+36895 r^2 z^2+9632 z^4\right\}
	-Q^6 \left\{[119701 M-49004 r] r^5+8 [21670 M-4053 r] r^3 z^2 \right. \\
&	\left. +r [32923 M+60726 r] z^4+44146 z^6\right\} +2025 M^2 \left\{r^2 [-2 M+r]^2-3 r [-2 M+r] z^2+z^4\right\} \rho ^6\\
&	+3 Q^4 \left\{2 M^2 \left[32985 r^6+44155 r^4 z^2+5033 r^2 z^4+775 z^6\right]-6 M r \left[8618 r^4-3695 r^2 z^2-7549 z^4\right] \rho ^2\right.\\
&	\left. +\left[9749 r^4-20300 r^2 z^2+2999 z^4\right] \rho ^4\right\} -3 Q^2 \left\{M^3 \left[37174 r^7+47910 r^5 z^2+5970 r^3 z^4+4450 r z^6\right]\right.\\
&	-M^2 \left[40349 r^6-20415 r^4 z^2-34905 r^2 z^4-2725 z^6\right] \rho ^2 +9 M r \left[1509 r^4-3090 r^2 z^2+295 z^4\right] \rho ^4\\
&	\left. -1350 \left[r^4-3 r^2 z^2+z^4\right] \rho ^6\right\} +675 a^4 \rho ^4 \left\{124 Q^4+15 M^2 \rho ^2+30 Q^2 \left[-5 M r+\rho ^2\right]\right\}\\
&	+675 a^2 \rho ^2 \left\{Q^6 \left[145 r^2+103 z^2\right]-2 Q^4 \left[(227 M-84 r) r^3+(143 M-30 r) r z^2+54 z^4\right] -15 M^2 \left[2 M r\right.\right.\\[-6pt]
&	\left.\left.-r^2+z^2\right] \rho ^4 +2 Q^2 \left[M^2 \left(7 r^2+z^2\right) \left(25 r^2+11 z^2\right)-2 M r \left(56 r^2-33 z^2\right) \rho ^2+15 \left(r^2-z^2\right) \rho ^4\right]\right\}\boldsymbol{\Big)} \Delta\tau^4 +O(\Delta\tau^6)\boldsymbol{\Bigg]}
\end{aligned}
\end{align}
\begin{align}
&ma^r =
	-\frac{q^2 \left(a^2+Q^2+r (-2 M+r)\right)}{6720 \Delta^2  \rho ^{18}} \boldsymbol{\Bigg[}112 Q^2 \rho^6 \Delta  \boldsymbol{\Big(}30 a^4 r+r \left\{Q^2+r [-2 M+r]\right\} \left\{5 Q^2+r [-8 M+3 r]\right\} +2 \left\{-2 M^2 r \right. \notag \\
&\hspace{0.6em}
\begin{aligned}
&	\left.-11 Q^2 r-9 r^3+M \left[Q^2+21 r^2\right]\right\} z^2+\{-4 M+9 r\} z^4+6 a^2 \left\{5 Q^2 r+4 r^3-6 r z^2+M \left[-9 r^2+z^2\right]\right\}\boldsymbol{\Big)}\Delta\tau^3 \\
&\hspace{-0.3em}	+\boldsymbol{\Big(}56 Q^{10} r \left\{28 r^2+z^2\right\}+14 Q^8 \left\{r^4 [-735 M+298 r]+10 [13 M-62 r] r^2 z^2+[M-270 r] z^4\right\} \\
&	+Q^6 \left\{r^5 \left[24955 M^2-19786 M r+3840 r^2\right]-2 r^3 \left[5285 M^2-23710 M r+8492 r^2\right] z^2+r \left[763 M^2+11622 M r\right. \right. \\
&	\left. \left.-4496 r^2\right] z^4+8 [-144 M+907 r] z^6\right\}+Q^4 \left\{r^6 \left[-26559 M^3+30642 M^2 r-11462 M r^2+1380 r^3\right]+r^4 \left[18645 M^3\right. \right.\\
&	\left.\left.-85118 M^2 r+56148 M r^2-9744 r^3\right] z^2-r^2 \left[3097 M^3+5010 M^2 r-4760 M r^2+1728 r^3\right] z^4+\left[83 M^3+1886 M^2 r \right.\right.\\
&	\left.\left.-25348 M r^2+9792 r^3\right] z^6+2 [607 M+198 r] z^8\right\}-9 M^2 \left\{[11 M-4 r] r^4 [-2 M+r]^2+3 r^2 \left[-4 M^3+80 M^2 r\right.\right.\\
&	\left.\left.-71 M r^2+16 r^3\right] z^2-9 r \left[4 M^2-19 M r+8 r^2\right] z^4+[-9 M+16 r] z^6\right\} \rho ^4+Q^2 \left\{4 M^4 r \left[2682 r^6-2531 r^4 z^2\right.\right.\\
&	\left.\left.+780 r^2 z^4-55 z^6\right]+M^2 r \left[8315 r^8-45844 r^6 z^2+4336 r^4 z^4+20872 r^2 z^6-1335 z^8\right]-2 M^3 \left[7865 r^8-25324 r^6 z^2\right.\right.\\
&	\left.\left.+2250 r^4 z^4-740 r^2 z^6+109 z^8\right]-6 M \left[309 r^8-2719 r^6 z^2+2639 r^4 z^4+41 r^2 z^6-26 z^8\right] \rho ^2+144 r \left[r^6-12 r^4 z^2\right.\right.\\
&	\left.\left.+18 r^2 z^4-4 z^6\right] \rho ^4\right\}+84 a^6 \rho ^2 \left\{176 Q^4 r+15 M^2 r \rho ^2+15 Q^2 \left[M \left(-15 r^2+z^2\right)+4 r \rho ^2\right]\right\}+2 a^2 \left\{56 Q^8 r \left[118 r^2\right.\right.\\
&	\left.\left.+37 z^2\right]-7 Q^6 \left[3 (1525 M-604 r) r^4+2 r^2 (299 M+588 r) z^2+(-89 M+1692 r) z^4\right]+2 Q^4 \left[3 r^5 \left(8491 M^2\right.\right.\right.\\
&	\left.\left.\left.-6487 M r+1202 r^2\right)+r^3 \left(-1918 M^2+16755 M r-5604 r^2\right) z^2+r \left(-175 M^2+16965 M r-7578 r^2\right) z^4\right.\right.\\
&	\left.\left.+3 (-369 M+544 r) z^6\right]+54 M^2 \left[r^3 \left(24 M^2-22 M r+5 r^2\right)+r \left(-4 M^2+45 M r-20 r^2\right) z^2+(-3 M+10 r) z^4\right] \rho ^4\right.\\
&	\left.+Q^2 \left[4 M^3 \left(-6812 r^6+1705 r^4 z^2-514 r^2 z^4+41 z^6\right)+6 M^2 r \left(4892 r^6-5365 r^4 z^2-3902 r^2 z^4+307 z^6\right)\right.\right.\\
&	\left.\left.+1080 \left(r^9-2 r^7 z^2-5 r^5 z^4+2 r z^8\right)-15 M \left(671 r^6-1839 r^4 z^2+399 r^2 z^4-3 z^6\right) \rho ^2\right]\right\}+6 a^4 \left\{168 Q^6 r \left[26 r^2+17 z^2\right]\right.\\
&	\left.+56 Q^4 \left[r^4 (-240 M+89 r)+2 r^2 (-61 M+13 r) z^2+(10 M-63 r) z^4\right]-45 M^2 \left[(13 M-6 r) r^2-(M-8 r) z^2\right] \rho ^4\right.\\
&	\left.+Q^2 \left[7 M^2 r \left(1459 r^4+550 r^2 z^2-45 z^4\right)-10 M \left(689 r^4-519 r^2 z^2+24 z^4\right) \rho ^2+360 r \left(3 r^2-4 z^2\right) \rho ^4\right]\right\}\boldsymbol{\Big)} \Delta\tau^5 \\
&	 + O(\Delta\tau^7) \boldsymbol{\Bigg]}
\end{aligned}
\end{align}
\begin{align}
&ma^\theta =
	\frac{a z \sin \theta }{6720 \Delta^2  \rho ^{18}} \boldsymbol{\Bigg[} 112 Q^2 \rho^6 \Delta \boldsymbol{\Big(}30 a^4+5 Q^4+r^2 \left\{20 M^2-28 M r+9 r^2\right\}+2 \left\{16 M-9 r\right\} r z^2+3 z^4\notag\\
&\hspace{0.6em}
\begin{aligned}
&	+6 a^2 \left\{5 Q^2-10 M r+6 r^2-4 z^2\right\}-2 Q^2 \left\{10 M r-7 r^2+8 z^2\right\}\boldsymbol{\Big)} \Delta\tau^3\\
&\hspace{-0.3em}	+ \boldsymbol{\Big(} 56 Q^{10} \left\{31 r^2+4 z^2\right\}-28 Q^8 \left\{5 [88 M-45 r] r^3+2 r [4 M+115 r] z^2+131 z^4\right\}+Q^6 \left\{r^4 \left[32515 M^2-32704 M r\right.\right.\\
&	\left.\left.+7912 r^2\right]-2 r^2 \left[1757 M^2-18928 M r+7508 r^2\right] z^2+\left[259 M^2+16128 M r-9752 r^2\right] z^4+4104 z^6\right\}\\
&	-2 Q^4 \left\{r^5 \left[18949 M^3-27967 M^2 r+13136 M r^2-1962 r^3\right]+r^3 \left[-4662 M^3+36765 M^2 r-26924 M r^2+4896 r^3\right] z^2\right.\\
&	\left.+r \left[581 M^3+10711 M^2 r-14632 M r^2+4536 r^3\right] z^4+3 \left[137 M^2+2428 M r-904 r^2\right] z^6-390 z^8\right\}\\
&	-18 M^2 \left\{[7 M-8 r] r^3 [-2 M+r]^2+18 r^2 \left[6 M^2-7 M r+2 r^2\right] z^2+3 [15 M-8 r] r z^4+2 z^6\right\} \rho ^4\\
&	+Q^2 \left\{28 M^4 r^2 \left[599 r^4-210 r^2 z^2+55 z^4\right]+8 M^3 r \left[-3999 r^6+5881 r^4 z^2+1135 r^2 z^4+327 z^6\right]+M^2 \left[21499 r^8\right.\right.\\
&	\left.\left.-47896 r^6 z^2-21452 r^4 z^4+12060 r^2 z^6+405 z^8\right]-120 M r \left[50 r^6-181 r^4 z^2+62 r^2 z^4+13 z^6\right] \rho ^2+144 \left[4 r^6-18 r^4 z^2\right.\right.\\
&	\left.\left.+12 r^2 z^4-z^6\right] \rho ^4\right\}+84 a^6 \rho ^2 \left\{176 Q^4+15 M^2 \rho ^2+60 Q^2 \left[-4 M r+\rho ^2\right]\right\}+4 a^2 \left\{28 Q^8 \left[127 r^2+46 z^2\right]\right.\\
&	\left.-14 Q^6 \left[(1328 M- 603 r) r^3+2 r (178 M+33 r) z^2+345 z^4\right]+27 M^2 \left[2 (7 M-5 r) (2 M-r) r^2+4 (9 M-5 r) r z^2\right.\right.\\
&	\left.\left.+5 z^4\right] \rho ^4-Q^2 \left[28 M^3 r \left(665 r^4+58 r^2 z^2+41 z^4\right)+3 M^2 \left(-7755 r^6+2152 r^4 z^2+4049 r^2 z^4+190 z^6\right)-540 \left(2 r^8\right.\right.\right.\\
&	\left.\left.\left.-5 r^4 z^4-2 r^2 z^6+z^8\right)+30 M r \left(303 r^4-416 r^2 z^2+9 z^4\right) \rho ^2\right]+Q^4 \left[7 M^2 \left(4583 r^4+750 r^2 z^2+55 z^4\right)\right.\right.\\
&	\left.\left.+42 M r \left(-671 r^4+130 r^2 z^2+369 z^4\right)+6 \left(992 r^4-1355 r^2 z^2+61 z^4\right) \rho ^2\right]\right\}+6 a^4 \left\{1512 Q^6 \left[3 r^2+2 z^2\right]\right.\\
&	\left.-224 Q^4 \left[(67 M-27 r) r^3+8 (5 M-2 r) r z^2+11 z^4\right]-90 M^2 \left[7 M r-4 r^2+3 z^2\right] \rho ^4\right.\\[-6pt]
&	\left.+Q^2 \left[-1120 M r \left(8 r^4+5 r^2 z^2-3 z^4\right)+7 M^2 \left(1755 r^4+950 r^2 z^2+59 z^4\right)+360 \left(4 r^2-3 z^2\right) \rho ^4\right]\right\}
\boldsymbol{\Big)}\Delta\tau^5 + O(\Delta\tau^7)	\boldsymbol{\Bigg]}
\end{aligned}
\end{align}
\begin{align}
&ma^\phi = 
	-\frac{q^2 a}{75600 \Delta ^{3/2} \rho ^{17}} \boldsymbol{\Bigg[} 
	11340 Q^2 \rho^8 \Delta \left(2 a^2+Q^2-2 M r+r^2-z^2\right) \Delta\tau^2 \notag \\
&\hspace{0.6em}
\begin{aligned}
&\hspace{-0.3em}	+\boldsymbol{\Big(} Q^8 \left\{23807 r^4+36895 r^2 z^2+9632 z^4\right\}-Q^6 \left\{[119701 M-49004 r] r^5+8 [21670 M-4053 r] r^3 z^2+r [32923 M \right.\\
&	\left.+60726 r] z^4+44146 z^6\right\}+2025 M^2 \left\{r^2 [-2 M+r]^2-3 r [-2 M+r] z^2+z^4\right\} \rho ^6+3 Q^4 \left\{2 M^2 \left[32985 r^6+44155 r^4 z^2\right.\right.\\
&	\left.\left.+5033 r^2 z^4+775 z^6\right]-6 M r \left[8618 r^4-3695 r^2 z^2-7549 z^4\right] \rho ^2+\left[9749 r^4-20300 r^2 z^2+2999 z^4\right] \rho ^4\right\}\\
&	-3 Q^2 \left\{M^3 \left[37174 r^7+47910 r^5 z^2+5970 r^3 z^4+4450 r z^6\right]-M^2 \left[40349 r^6-20415 r^4 z^2-34905 r^2 z^4-2725 z^6\right] \rho ^2\right.\\
&	\left.+9 M r \left[1509 r^4-3090 r^2 z^2+295 z^4\right] \rho ^4-1350 \left[r^4-3 r^2 z^2+z^4\right] \rho ^6\right\}+675 a^4 \rho ^4 \left\{124 Q^4+15 M^2 \rho ^2 + 30 Q^2 \left[-5 M r\right.\right.\\
&	\left.\left.+\rho ^2\right]\right\}+675 a^2 \rho ^2 \left\{Q^6 \left[145 r^2+103 z^2\right]-2 Q^4 \left[(227 M-84 r) r^3+(143 M-30 r) r z^2+54 z^4\right] -15 M^2 \left[2 M r\right.\right.\\
&	\left.\left.-r^2+z^2\right] \rho ^4+2 Q^2 \left[M^2 \left(7 r^2+z^2\right) \left(25 r^2+11 z^2\right)-2 M r \left(56 r^2-33 z^2\right) \rho ^2+15 \left(r^2-z^2\right) \rho ^4\right]\right\} \boldsymbol{\Big)}\Delta\tau^4+ O(\Delta\tau^6)\boldsymbol{\Bigg]}
\end{aligned}
\end{align}
\end{subequations}
As expected, in the limit $a \rightarrow 0$, these reduce to Eqs.~(\ref{eq:staticresult}).

%---------------------------------------------------------------------------------------------------------

\section{Conclusions}
Our previous work in Paper~I was only valid provided the particle's path is a geodesic of the background spacetime. We have shown here how this may be extended with reasonable ease to allow for non-geodesic motion. Some interesting insights have come as a result of this study.

First, it is interesting to note that, although the covariant approach still has the great benefit of being sufficiently general to be applicable to any spacetime of interest, we must convert to a coordinate expansion in order to obtain a final result. This is not very suprising as the benefit of using a covariant expansion hinges on having a geodesic along which to expand. The coordinate expansion, on the other hand, more naturally characterises motion which is not along a geodesic. However, as Sec.~\ref{sec:kn} demonstrates, beginning with the covariant expansion and computing the coordinate expansion from it allows the calculation to be done for space-times which previous methods have not yet been able to do.

It is also interesting to look more closely at the presence of the Christoffel symbols in the relation between the covariant and coordinate expansions. In the present context, these take on a particularly clear form by being directly related to the 4-acceleration of the particle. In particular, the leading order at which this effect appears is given in Eq.~(\ref{eq:vab-hat}), which in this case becomes (since the first two orders, $\hat{v}$ and $\hat{v}_a$, are identically zero):
\begin{equation}
\hat{v}_{ttt}	= v_{ttt} + 3 v_{t r} \Gamma_{t t}^{r} = v_{ttt} + 3 v_{t r} a^{r}
\end{equation}
where $a^r$ is the radial component of the 4-acceleration of the particle. It is immediately clear from this expression that the deviation from the result for geodesic motion is a direct consequence of the particle's 4-acceleration.

%---------------------------------------------------------------------------------------------------------

\section{Acknowledgments}
BW is supported by the Irish Research Council for Science, Engineering and Technology, funded by the National
Development Plan. 

We would like to thank Paul Anderson and Ardeshir Eftekharzadeh for the considerable time and
effort they gave to resolving some issues that arose during this work. We would also like to thank Warren Anderson
and Antoine Folacci whose ideas and suggestions were invaluable. Also our appreciation to the many others from
the Capra meetings for interesting and useful conversations. Finally, we would like to thank Kirill Ignatiev for discovering the relation given in Eq.~(\ref{eq:vcompsym2}) and also thank Marc Casals and Sam Dolan for their continuous feedback and discussions.

%---------------------------------------------------------------------------------------------------------

%%%%%%%%%%%%%%%%%%%%%%%%%%%%%%%%%%%%%%%%%%%%%%%%%%%%%%%%%%%%%%%%%%%%%%%%%%%%%%%%%%%%%%%

\bibliography{references}{}
\bibliographystyle{apsrev}

\end{document}